\documentclass[a4paper,12pt]{article}

\usepackage{epsfig,wrapfig}

\begin{document}
\setlength{\unitlength}{1mm}

\begin{titlepage}

\begin{flushright}
Edinburgh 2003/06\\
IPPP/03/28\\
DCPT/03/56\\
May 2003
\end{flushright}
\vspace{1.cm}

\begin{center}
\large\bf

{\LARGE\bf Numerical evaluation of multi-loop integrals by sector decomposition}\\[2cm]

\rm
{T.~Binoth$^{a}$ and G.~Heinrich$^{b}$ }\\[.5cm]

{\em $^{a}$School of Physics, The University of Edinburgh,\\
	   EH9 3JZ Edinburgh, Scotland, UK} \\[.5cm]

{\em $^{b}$IPPP, University of Durham, Durham DH1 3LE, UK} \\[.5cm]

\end{center}
\normalsize

\begin{abstract}
In a recent paper\,\cite{Binoth:2000ps} we have presented an automated
subtraction method for divergent multi-loop/leg integrals
in dimensional regularisation which allows for their numerical
evaluation, and applied it to diagrams 
with massless internal lines.
Here we show how to extend this algorithm to 
Feynman diagrams with massive propagators and arbitrary propagator powers.  
As applications, we present numerical results for the 
master 2-loop 4-point topologies with massive internal lines
occurring in Bhabha scattering at two loops, and for the 
master integrals of planar and non-planar 
massless double box graphs with two off-shell legs. 
We also evaluate numerically some 
two-point functions up to 5 loops relevant for  beta-function calculations,   
and a 3-loop 4-point function, the massless on-shell planar triple box.
Whereas the 4-point functions are evaluated in non-physical kinematic regions, 
the results for the propagator functions are
valid for arbitrary kinematics.  
\end{abstract}

\vspace{3cm}

\end{titlepage}

\section{Introduction}
The precision of present and upcoming high energy experiments is unprecedented 
and clearly requires an analogous level of accuracy on the theoretical side. 
One of the steps towards achieving this goal is 
to push the level of higher order corrections in perturbation theory further. 
However, the calculation of the corresponding Feynman diagrams becomes more and more 
involved as the number of loops and legs of the graphs increases. 
Typically the occurring tensor integrals are expressed in terms of scalar integrals which 
are then  reduced to a set of master topologies. These master topologies  
are the basic ingredients and thus the bottleneck for
any attempt to perform a higher order calculation.   
 
A milestone in this respect has been the analytic calculation of the master 
integrals for the 
planar~\cite{Smirnov:1999gc} and non-planar~\cite{Tausk:1999vh} massless two-loop box diagrams 
with on-shell external legs and the ones with one external leg 
off-shell~\cite{Smirnov:2000vy,Gehrmann:2000zt,Smirnov:2000ie,Gehrmann:2001ck}. 
These results triggered the analytical computation of two-loop matrix elements 
for 2 $\to$ 2 scattering processes with on-shell 
legs~\cite{Bern:2000ie}
and 1 $\to$ 3  processes with one off-shell leg~\cite{Garland:2001tf}, 
which are necessary for the next-to-next-to-leading order 
calculation of prominent processes like 
two-jet production in hadronic collisions or three-jet production
in $e^+e^-$ collisions.
On the other hand, the two-loop four-point functions with two external 
legs off-shell still await their analytical calculation. 
These integrals are for example needed to calculate the production of two 
massive vector bosons in hadronic collisions at two loops. 

The reason for the lack of analytical results is the 
increasing number of scales as the number of off-shell legs increases, 
leading to more complicated analytic structures. 
The same argument is true if the diagrams contain massive 
internal lines. Such diagrams are certainly  important  
in view of a forthcoming  $e^+e^-$ collider at TeV energies, 
where the experimental precision
will be such that the Standard Model is tested at the two-loop level.
However, a major part of the corresponding two-loop master integrals
with massive propagators may not be accessible analytically.
Thus the development of numerical methods seems natural to make progress
in loop calculations. 

This is not at all straightforward,  
as different kinds of singularities -- infrared (IR), ultraviolet (UV)
and threshold singularities -- are present.  
Even in the absence of IR and UV divergences, one has to face problems due to  
a complicated analytic structure:
For physical kinematics the integral generically has singularities 
within the integration region, which typically hinder a successful, 
numerically stable evaluation. 
At the one-loop level this problem is less severe 
because the integration space can be reduced to lower dimensionality
and because the analytical structure is completely understood. 
Solutions  for multi-leg one-loop diagrams have been suggested 
in~\cite{Ferroglia:2002mz,Binoth:2002xh}. However, beyond one loop 
the problem is not satisfactorily solved in a general and constructive way, 
although promising approaches have been proposed~\cite{
Tkachov:1996wh,Ghinculov:2000cz,Passarino:2001jd,smirnovbuchneu}.
In the presence of infrared divergences the situation is even more difficult.
One first has to obtain a numerically integrable function by 
subtracting the infrared poles. As these poles are in general 
overlapping, this is a highly non-trivial task. 
In this paper  we will focus on this problem, i.e. on the isolation of infrared 
divergences from general multi-loop Feynman diagrams. 

In~\cite{Binoth:2000ps} we proposed a constructive method which enables to 
calculate multi-loop integrals numerically for unphysical kinematics, 
and applied it to diagrams with massless propagators. 
It served to check the results for the massless
on-shell double boxes and to make predictions for the massless 
doubles boxes with one off-shell leg, the latter being confirmed by 
the subsequent analytical calculation later. 
This method is based on {\it sector decomposition}, 
originally used to disentangle overlapping UV divergences in the proof 
of the BPHZ theorem~\cite{hepp}. 
The basic concept being very simple, it  has a wide range of applicability. 
For example, 
it  has been used 
to derive high energy approximations for one-loop integrals~\cite{roth}. 
Sector decomposition also has been employed to 
extract logarithmic mass singularities from massive multi-scale integrals 
in the high energy limit at two loops~\cite{Denner:2003wi}. 
In~\cite{Binoth:2000ps}, the concept of sector decomposition has been elaborated 
to a highly automated program package, allowing to convert  
dimensionally regulated Feynman integrals 
into a Laurent series in $\epsilon$. 
The coefficients of the poles are sums of multi-dimensional parameter 
integrals. 
In a kinematic region where thresholds are absent,   
the latter can be evaluated with standard numerical integration routines.   
In simple cases, analytic integration is also possible. 
The numbers of loops and legs are
in principle arbitrary, but 
of course limited by CPU time and disk space.  
Recently the method has been extended to deal also with 
phase space integrals~\cite{radcor}. 

In this article, we show how to generalise the proposed method
to arbitrary Feynman diagrams. This is achieved
by extending the algorithm to graphs with arbitrary  masses 
and propagator exponents.
With these extensions one is able to deal with very different types
of multi-loop/multi-scale problems, most of which are at the edge or
beyond the present state of the art of analytic calculations. 
More precisely, we present numerical results for massless two-loop 
4-point functions with two external legs off-shell, 
and for the three most difficult master topologies for Bhabha scattering 
at two loops, at two numerical points where $s,t<0$. 
We also give numerical results for 3-loop, 4-loop and 5-loop 
massless 2-point functions
relevant for the calculation of beta-functions, and 
finally for the massless  on-shell planar 3-loop box diagram. 
Although we do not provide a solution for arbitrary physical kinematics
in general, we note that there is an important  class of diagrams,
namely diagrams with only one scale, where there is no kinematical restriction.  

The paper is organised as follows. In section 2, we 
briefly present the method, extended to include also non-integer 
propagator powers, respectively Feynman parameters in the numerator. 
In section 3 we collect applications of the method, presenting 
numerical results for the  types of diagrams listed above. 
Section 4 contains the discussion of our results and the conclusions.

\section{The formalism} 

The  formalism of iterated sector decomposition 
enables to extract in a constructive way the 
divergences of arbitrary scalar $L$-loop $N$-point integrals, 
where the divergences are regularised by dimensional regularisation. 
The method has been described in detail in~\cite{Binoth:2000ps}. Here 
we only sketch it briefly, focusing on the features that 
have not been elaborated previously, i.e. the possibility
of arbitrary propagator powers and the inclusion of 
massive propagators. 

A scalar graph $G$ with  $N$ propagators and $L$ 
$D$-dimensional loop momenta, where 
the propagators can have arbitrary, not necessarily integer powers $\nu_j$,  
has the following representation in momentum and Feynman parameter
space\footnote{This representation, as well as the following discussion 
of its topological properties, is well known, see for 
example~\cite{nakanishi,zavialov,itzyksonzuber,tarasov,smirnovbuchneu}.}:
\begin{eqnarray}\label{eq0}
G  &=&
\int \prod\limits_{l=1}^{L} \frac{d^Dk_l}{i\pi^{\frac{D}{2}}}\;
\prod\limits_{j=1}^{N} \frac{1}{(q_j^2-m_j^2+i\delta)^{\nu_j}}
\nonumber\\
&=&(-1)^{N_{\nu}}\frac{\Gamma(N_{\nu}-LD/2)}{\prod_{j=1}^{N}\Gamma(\nu_j)}
\int\limits_{0}^{\infty} \prod\limits_{j=1}^{N}\,dx_j\,x_j^{\nu_j-1} 
\delta(1-\sum_{i=1}^N x_i)\,\frac{{\cal U}^{N_{\nu}-(L+1) D/2}}
{{\cal F}^{N_{\nu}-L D/2}}\label{eq1}
\end{eqnarray}
Here $q_j$ are the propagator momenta, i.e. linear combinations of external and loop
momenta, $N_{\nu}=\sum_{j=1}^N\nu_j$. The functions ${\cal U}$ and ${\cal F}$ can be 
straightforwardly derived from the momentum representation. 
They also can be constructed
from the topology of the corresponding Feynman graph as follows. 

Cutting $L$ lines of a given connected $L$-loop graph such that it becomes a connected
tree graph $T$ defines a {\em chord} ${\cal C}(T)$ as being the set of lines 
not belonging to this tree. The Feynman parameters associated with each chord 
define a monomial of degree $L$. The set of all such trees (or {\em 1-trees}) 
is denoted by ${\cal T}_1$.  The 1-trees $T\in {\cal T}_1$ define 
${\cal U}$ as being the sum over all monomials corresponding 
to a chord ${\cal C}(T\in {\cal T}_1)$.
Cutting one more line of a 1-tree leads to two disconnected trees, or a  {\em 2-tree} $\hat T$.
${\cal T}_2$ is the set of all such  2-trees.
The corresponding chords define  monomials of degree $L+1$. Each 2-tree of a graph
corresponds to a cut defined by cutting the lines which connected the 2 now disconnected trees
in the original graph. The momentum flow through the lines of such a cut defines a Lorentz invariant
$s_{\hat T} = ( \sum_{j\in \rm Cut(\hat T)} p_j )^2$.   
The function ${\cal F}_0$ is the sum over all such monomials times 
minus the corresponding invariant:    
\begin{eqnarray}\label{eq0def}	
{\cal U}(\vec x) &=& \sum\limits_{T\in {\cal T}_1} \Bigl[\prod\limits_{j\in {\cal C}(T)}x_j\Bigr]\;,\nonumber\\
{\cal F}_0(\vec x) &=& \sum\limits_{\hat T\in {\cal T}_2}\;
\Bigl[ \prod\limits_{j\in {\cal C}(\hat T)} x_j \Bigr]\, (-s_{\hat T})\;,\nonumber\\
{\cal F}(\vec x) &=&  {\cal F}_0(\vec x) + {\cal U}(\vec x) \sum\limits_{j=1}^{N} x_j m_j^2\;.
\end{eqnarray}  
${\cal U}$ is a positive semi-definite function. 
Its vanishing is related to the  UV subdivergences of the graph. 
Overall UV divergences, if present,
will always be contained in the  prefactor $\Gamma(N_{\nu}-LD/2)$. 
In the Euclidean region, ${\cal F}$ is also a positive semi-definite function 
of the Feynman parameters $x_j$.  Its vanishing does not necessarily lead to 
an IR singularity. Only if some of the invariants are zero, 
for example if some of the external momenta
are light-like, the vanishing of  ${\cal F}$  may induce an IR divergence.
Thus it depends on the {\em kinematics}
and not only on the topology (like in the UV case) 
whether a zero of ${\cal F}$ leads to a divergence or not. 
This fact makes it much harder to formulate
general theorems on the IR singularity structure of Feynman graphs.
Examples of ${\cal U}$ and ${\cal F}$ functions 
will  be given  below. 

In multi-loop integrals, the singular regions -- 
leading to at most $1/\epsilon^{2L}$ infrared poles upon 
integration in parameter space --
are generally {\it overlapping}, such that the set-up of a simple subtraction
scheme  becomes impossible. In~\cite{Binoth:2000ps} we have developed
an algorithm and computer program 
to disentangle such overlapping regions. 
It can briefly be sketched as follows: 
\begin{description}
\item[Step 1:]
A  "primary" sector decomposition of all Feynman parameters
eliminates the $\delta$-function and maps the integral 
to a sum of $N$ ($N$--1)-dimensional parameter integrals
over the unit cube: The decomposition 
$$
\int d^N x =
\sum\limits_{l=1}^{N} \int d^N x
\prod\limits_{\stackrel{j=1}{j\ne l}}^{N}\theta(x_l\ge x_j\ge 0)$$
and appropriate variable substitutions lead to
$$G= \sum\limits_{l=1}^{N}G_l\,, \;
G_l = \int\limits_{0}^{1} \prod\limits_{i=1}^{N-1}dx_i \,
\frac{ {\cal U}_l^{N-(L+1)D/2}(\vec{x})}{ {\cal F}_l^{N-L D/2}(\vec{x})}\;,
$$ 
where ${\cal U}_l, {\cal F}_l$ contain 
only positive semi-definite functions and where the singularities
can only be located at $x_i=0$.
\item[Step 2:]
Sector decomposition is performed iteratively for sets of parameters $\{x_k\}$
which make the transformed $\hat{\cal F}$ or $\hat{\cal U}$ functions vanish
at $x_k=0$. 
The iteration stops 
if  $\hat{\cal U}=1+ \dots$ and  $\hat{\cal F}= (-s_{ij}) + \dots$ , where
the dots denote positive semi-definite remainder terms and $(-s_{ij})$ is a 
kinematic invariant which is assumed to be positive. At this point all divergences are
non-overlapping, the poles will be entirely contained in overall factors 
$x_i^{\alpha+\beta\,\epsilon}$, with $\alpha$ a negative integer.
\item[Step 3:]
Subtractions are per\-formed  using Tay\-lor expan\-sion 
around $x_i=0$. After subtraction, it is safe to 
expand in $\epsilon$ up to the desired order, leading to the form 
\begin{equation}
G=\sum\limits_{j=-M}^{2L}\frac{C_j(\vec{x})}{\epsilon^j} + O(\epsilon^{M+1})\; .
\label{laurent}
\end{equation}
Note that the method 
in principle allows to calculate coefficients of the Laurent series 
in $\epsilon$ to arbitrary order $M$.
\item[Step 4:]    
The coefficients $C_m(\vec{x})$ in (\ref{laurent}) are sums 
of subtracted and expanded  
subsector integrals in terms of bounded functions of Feynman parameters. 
These can be evaluated numerically, or even analytically in simple cases.
\end{description}   

In~\cite{Binoth:2000ps} we were concentrating on massless integrals.
The iteration always stopped, as each sector decomposition 
reduces the degree of the monomials which are relevant for the
IR behaviour of the integral. 
Including propagator masses, it is clear that the
IR behaviour of a diagram can only improve, and hence this  
extension of the algorithm does not pose any principle problem. 
On the other hand, a new technical feature appears, which is that 
the Feynman parameters associated with massive lines occur  
quadratically in ${\cal F}$, stemming from the term proportional to
${\cal U}$ in eq.~(\ref{eq0def}).  
In this case it is not guaranteed 
that the iteration stops if one proceeds thoughtlessly. 
For example, it is easy to see that a polynomial of the type $x\,y^2+z^2$ 
can produce an endless
loop if a sector decomposition in $\{x,z\}$ is followed by one in $\{x,y\}$.
A possible way out is to decompose first all parameters which occur
quadratically, in our example $\{y,z\}$. 
Then no dangerous situation is present anymore and one can proceed as usual. 
The corresponding modification of the code is straightforward, and 
examples of such a situation are the two-loop QED graphs calculated below.    

We note that with the generalisation of the algorithm presented in this article,
integrals with arbitrary numerators, and thus tensor integrals, can be treated. 
As is well known, after integration over the loop momenta, tensor integrals 
correspond to linear combinations of integrals with monomials of Feynman parameters in 
the numerator (times tensors  carrying the Lorentz structure). 
These numerators are taken into account by our program 
and  may cancel IR singularities 
which would be present in the scalar diagram. 
Another possibility is to express 
tensor integrals by  Feynman
parameter integrals which do not have polynomial numerators, 
but are partly in shifted dimensions, 
as advocated in\,\cite{Davydychev:1991va,tarasov}.
As our method treats the space-time dimension as a parameter, 
this way is viable as well. 


\section{Applications}

In this section we discuss a number of applications of our algorithm, 
ranging from 2-loop 4-point functions to 5-loop 2-point functions.
We note that sector decomposition also is useful for the calculation 
of graphs which are not plagued by IR or UV divergences, 
as it produces integrands which are bounded and positive definite in  
kinematic regions where no thresholds are approached. 
The technical problem of integrable singularities inside multi-dimensional
integration regions hinders the direct computation of the parameter integrals
for general kinematics. However, this  is not a limitation intrinsic to 
our procedure, as more sophisticated numerical integration routines than the
one we use could overcome this problem. However, for the time being, 
we have to restrict ourselves to the non-physical region, where all 
Mandelstam variables are negative, in the case of the box functions.  

For the numerical evaluations we  have used the standard  
Monte Carlo package BASES \cite{kawabata}.
The numerical precision  for all quoted results  
is 1\% or better.
 
The integration time for all 2-loop boxes  to reach the demanded 
precision is of the order of an hour on a PC with a Pentium IV 
(2\,GHz) processor. For the 3-loop propagators the required time is less, 
whereas for the 5-loop propagator it took a few days to reach this precision. 

\subsection{Parameter representation of general double box graphs}

As the topological functions ${\cal U}$
and ${\cal F}$ are the basic input for our algorithm,  
we first give the parameter representation of the general planar 
and non-planar double box topologies. These representations 
are used as a starting point for the particular 2-loop box 
examples calculated below.

\subsection*{Planar topology}

The functions ${\cal U}$ and ${\cal F}$ for the  
planar double box graph $G_{\rm{P}}(s,t,u,s_1,s_2,s_3,s_4,\{m_j^2\})$, 
as shown in Fig.\,(\ref{figp}a), are given by
\begin{eqnarray}
{\cal U} &=& x_{123} x_{567} + x_{4} x_{123567} \nonumber\\
{\cal F} &=& \quad(-s)\, (x_2 x_3 x_{4567} + x_5 x_6 x_{1234} 
                         + x_2 x_4 x_6 + x_3 x_4 x_5)   
			 +\,(-t)\, x_1 x_4 x_7  \nonumber\\
 && +\, (-s_1)\, x_1 \,( x_4 x_5 + x_2 x_{4567} )  
    + \,(-s_2)\, x_1\, ( x_4 x_6 + x_3 x_{4567} ) \nonumber\\
 && +\, (-s_3)\, x_7\, ( x_3 x_4 + x_6 x_{1234} ) 
    + \,(-s_4)\, x_7\, ( x_2 x_4 + x_5 x_{1234} ) \nonumber\\
	  && + \,{\cal U}\,\sum\limits_{j=1}^{7} x_j \,m_j^2
	  \label{eqplanar}
\end{eqnarray}  
where we use the short-hand notations $s_i=p_i^2$, $s=(p_1+p_2)^2,  
t=(p_2+p_3)^2$ and 
$x_{i_1\ldots i_n}=x_{i_1}+\ldots+x_{i_n}$. 
In the examples considered in the following, we will set some of the
$s_i$ and/or $m_j^2$ to zero. 

\subsection*{Non-planar topology}

\begin{figure}[ht]
\begin{picture}(100,35)
\put(5,22){(a)}
\put(10,0){\epsfig{file=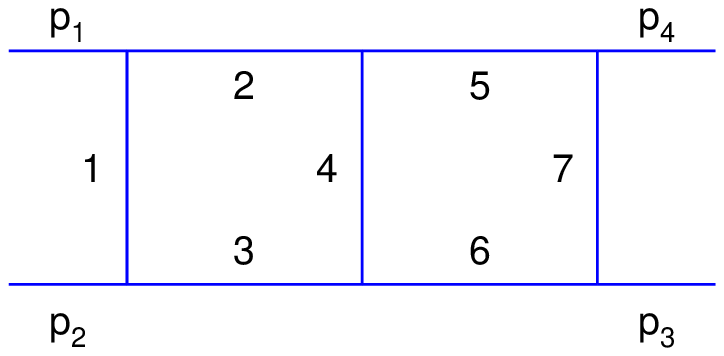,height=3.cm}}
\put(85,22){(b)}
\put(90,0){\epsfig{file=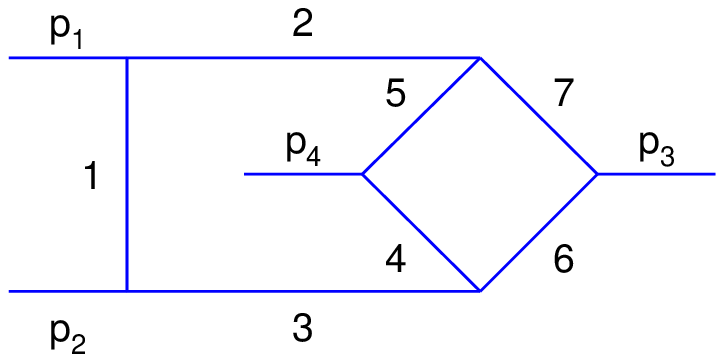,height=3.cm}}
\end{picture}
\caption{\em The (a) planar and (b) non-planar double box.}\label{figp}
\end{figure}

The parameter representation  of the general non-planar 
double box \\$G_{\rm{NP}}(s,t,u,s_1,s_2,s_3,s_4,\{m_j^2\})$, 
Fig.\,(\ref{figp}b), is given by
\begin{eqnarray}
{\cal U}&=&x_{123}x_{4567}+x_{45}x_{67}\nonumber\\
{\cal F}&=&(-s)\,(x_2x_3x_{4567}+x_2x_4x_6+x_3x_5x_7)\nonumber\\
	      &&+\,(-t)\,x_1x_4x_7+(-u)\,x_1x_5x_6\nonumber\\
	      &&+\,(-s_1)\,x_1(x_5x_7+x_2x_{4567})
	      +\,(-s_2)\,x_1(x_4x_6+x_3x_{4567})\nonumber\\
              &&+\,(-s_3)\,(x_6x_7x_{12345}+
                             x_3x_4x_7+x_2x_5x_6)\nonumber\\
              &&+\,(-s_4)\,(x_4x_5x_{12367}+
                   x_3x_5x_6+x_2x_4x_7)\nonumber\\
	  && + \, {\cal U}\,\sum\limits_{j=1}^{7} x_j \,m_j^2\;.
	  \label{eqnonplanar}
\end{eqnarray}

\subsection{Massless double box graphs with two legs off-shell}

For the massless 4-point functions  considered in this subsection, 
two legs are off-shell and  the internal 
lines are massless.
The topologically different positions of the off-shell legs give rise to 
three master integrals for both the planar- and the non-planar double box
with two legs off-shell.
We give results at two Euclidean numerical points for each master integral, as shown 
in Tables~\ref{planar} and \ref{nonplanar}. 
An overall factor $\Gamma^2(1+\epsilon)$ has been extracted: 
$$G_{\rm{P},\rm{NP}}(s,t,u,s_1,s_2,s_3,s_4) = 
\Gamma^2(1+\epsilon)\sum_{i=0}^4 \frac{P_i}{\epsilon^i}$$.

\begin{table}[ht]
\begin{center}
\begin{tabular}{|c||c|c|c|}
\hline
\multicolumn{4}{|c|}{}\\
\multicolumn{4}{|c|}{$(-s,-t,-u)=(2/3,2/3,2/3)$}\\
\multicolumn{4}{|c|}{}\\
\hline
&&&\\
$(-s_1,-s_2,-s_3,-s_4)$&$(1,0,0,1)$&$(0,1,0,1)$&$(0,0,1,1)$\\
&&&\\
\hline
&&&\\
$P_4$&-0.8437 &0     &-0.8438\\
$P_3$&-2.052  &0     &-2.053\\
$P_2$&-2.189  &-3.571&-10.53\\
$P_1$&4.394   &-6.585&-48.60\\
$P_0$&35.64   &11.83 &-140.7\\
&&&\\
\hline
\hline
\multicolumn{4}{|c|}{}\\
\multicolumn{4}{|c|}{$(-s,-t,-u)=(1/2,1/3,5/6)$}\\
\multicolumn{4}{|c|}{}\\
\hline
&&&\\
$(-s_1,-s_2,-s_3,-s_4)$&$(2/3,0,0,1)$&$(0,2/3,0,1)$&$(0,0,2/3,1)$\\
&&&\\
\hline
&&&\\
$P_4$&-3.000 &0     &-3.000\\
$P_3$&-12.47 &0     &-14.91\\
$P_2$&-26.28 &-7.827&-73.79\\
$P_1$&-19.04 &-18.90&-325.3\\
$P_0$&90.68  &16.18 &-1151.\\
&&&\\
\hline
\end{tabular}
\end{center}
\caption{\em Numerical results for the planar double box with two off-shell legs}
\label{planar}
\end{table}

We would like to note at this point that our method is not a priori 
a numerical method. In principle, the functions obtained after step 3 
of the algorithm
can be integrated analytically. However, as the parameter integrals 
for the coefficient of the $1/\epsilon^l$ pole are $(N-1-l)$--dimensional,
and the functions to integrate are in general highly non-trivial, 
only the integrals for large $l$, i.e. the highest pole coefficients,  
can be evaluated automatically by standard algebraic integration
packages.   
For example,  one obtains the following analytical result for the 
leading and subleading poles of the planar 
topology with $s_3\not = 0, s_4\not = 0$\,:
\begin{eqnarray*}
G_{\rm{P}}(s,t,u,0,0,s_3,s_4)&=&\Gamma^2(1+\epsilon)\,
(-s)^{-2\epsilon}\frac{1}{s^2 t}
\Big\{\;\frac{1}{4\epsilon^4} \\
&&+ \frac{1}{\epsilon^3}
\Big[\,-\log(-t)+\frac{1}{2}\log(-s_3)+\frac{1}{2}\log(-s_4)
\Big] \Big\}\;+ \;{\cal O}(\frac{1}{\epsilon^2})
\end{eqnarray*}
Although the analytic integration could be pushed further by 
writing algebraic integration routines which are specialised to 
deal with the functions occurring in these integrals, we do not 
pursue the analytic approach further here. 

\begin{table}[ht]
\begin{center}
\begin{tabular}{|c||c|c|c|}
\hline
\multicolumn{4}{|c|}{}\\
\multicolumn{4}{|c|}{$(-s,-t,-u)=(2/3,2/3,2/3)$}\\
\multicolumn{4}{|c|}{}\\
\hline
&&&\\
$(-s_1,-s_2,-s_3,-s_4)$&$(1,1,0,0)$&$(0,1,0,1)$&$(0,0,1,1)$\\
&&&\\
\hline
&&&\\
$P_4$&-2.812 &-0.5625 &0\\
$P_3$&-3.192 &-0.0923&2.189\\
$P_2$&29.82  &5.481   &1.634\\
$P_1$&139.2  &35.40   &-17.98\\
$P_0$&365.3  &170.6   &-60.33\\
&&&\\
\hline
\hline
\multicolumn{4}{|c|}{}\\
\multicolumn{4}{|c|}{$(-s,-t,-u)=(1/2,1/3,5/6)$}\\
\multicolumn{4}{|c|}{}\\
\hline
&&&\\
$(-s_1,-s_2,-s_3,-s_4)$&$(2/3,1,0,0)$&$(0,2/3,0,1)$&$(0,0,2/3,1)$\\
&&&\\
\hline
&&&\\
$P_4$&-6.700 &-0.600&0\\
$P_3$&-15.97 &1.377 &4.502\\
$P_2$&38.24  &11.79 &5.870\\
$P_1$&305.1  &57.33 &-36.18\\
$P_0$&1080.  &267.0 &-165.5\\
&&&\\
\hline
\end{tabular}
\end{center}
\caption{\em Numerical results for the non-planar double box with two off-shell legs}
\label{nonplanar}
\end{table}

\subsection{Double Box graphs for Bhabha scattering}

Bhabha scattering is a paradigm process of QED. Its calculation 
at the two-loop level, apart from the theoretical challenge alone, 
has a strong motivation by the fact that Bhabha scattering serves as
a luminosity monitor for $e^+e^-$ colliders.

The virtual two-loop QED corrections to Bhabha scattering have been 
calculated analytically by Bern, Dixon and Ghinculov~\cite{Bern:2000ie}, 
in the limit of vanishing electron mass. Here we give 
numerical  results for the three 2-loop box master integrals 
entering these corrections with {\it massive} internal lines. 
The numerical points are calculated 
for nonphysical kinematics, $s,t<0$, but they can serve as a strong check 
for a future analytic calculation which includes the fermion masses.

\begin{figure}[ht]
\begin{picture}(100,25)
\put(2,9){(a)}
\put(5,0){\epsfig{file=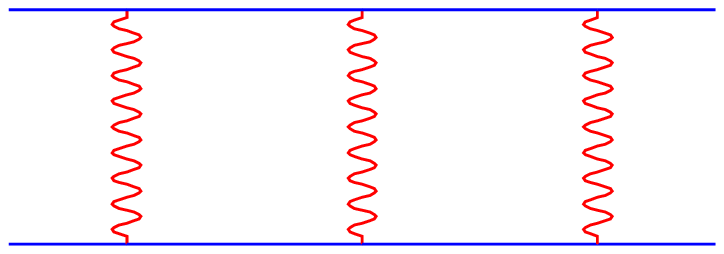,height=2.cm}}
\put(55,9){(b)}
\put(58,0){\epsfig{file=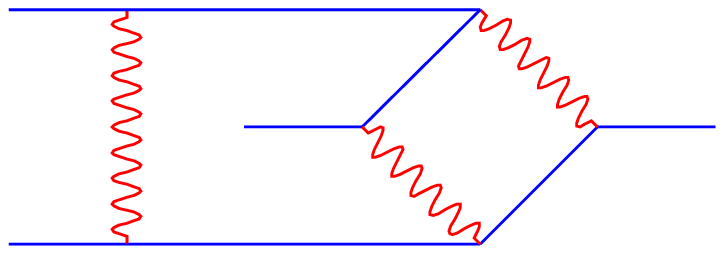,height=2cm}}
\put(104,9){(c)}
\put(105,0){\epsfig{file=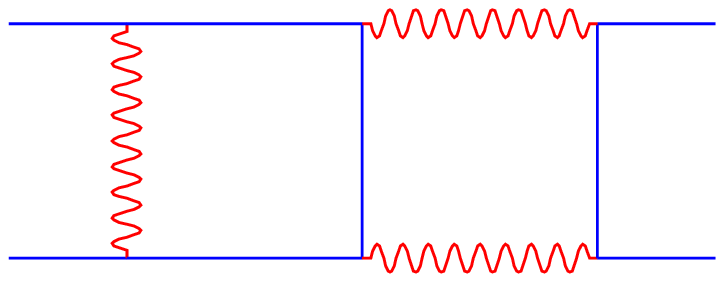,height=2cm}}
\end{picture}
\caption{\em The two-loop four point master topologies relevant for Bhabha scattering. The wavy lines
are massless (photons) and the straight lines are massive scalars with external 
legs on-shell.
We label the topologies from left to right by $G_a$, $G_b$, $G_c$.}
\label{bhabha_graphs}
\end{figure}

The three master topologies $G_{l=a,b,c}$ are shown in Fig.~\ref{bhabha_graphs}. 
The corresponding functions ${\cal U}$ and ${\cal F}$ are given by 
equation (\ref{eqplanar}) for
$G_a$ and $G_c$ and by  (\ref{eqnonplanar}) for $G_b$:
\begin{eqnarray}
G_a(s,t,u,m^2,M^2) &=& \;\;G_{\rm{P}}(s,t,u,m^2,M^2,M^2,m^2;0,m^2,M^2,0,m^2,M^2,0)\nonumber\\
G_b(s,t,u,m^2,M^2) &=& G_{\rm{NP}}(s,t,u,m^2,M^2,M^2,m^2;0,m^2,M^2,0,m^2,M^2,0)\nonumber\\
G_c(s,t,u,m^2,M^2) &=& \;\;G_{\rm{P}}(s,t,u,m^2,m^2,M^2,M^2;0,m^2,m^2,m^2,0,0,M^2)\nonumber
\end{eqnarray}
Results for two different numerical points are given in 
Table~\ref{tablebhabha}. 
An overall factor $\Gamma^2(1+\epsilon)$ has been extracted:
$$G_{l=a,b,c}(s,t,u,m^2,M^2) = 
\Gamma^2(1+\epsilon)\sum_{i=0}^2 \frac{P_i}{\epsilon^i}\;.$$
To be slightly more general, the second  point covers the case 
where the two massive propagator lines flowing through
the graphs have different masses. 

\small
\begin{table}[ht]
\begin{center}
\begin{tabular}{|c||c|c|c||c|c|c|}
\hline
&\multicolumn{3}{|c||}{}&\multicolumn{3}{|c|}{}\\
$(-s,-t,-u,m^2,M^2)$&\multicolumn{3}{|c||}{$(1/5,3/10,7/2,1,1)$}&
\multicolumn{3}{|c|}{$(5/3,4/3,5,1,3)$}\\
&\multicolumn{3}{|c||}{}&\multicolumn{3}{|c|}{}\\
\hline
&&&&&&\\
&$G_a$&$G_b$&$G_c$&$G_a$&$G_b$&$G_c$\\
&&&&&&\\
\hline
&&&&&&\\
$P_2$&-1.561   &-0.5255 &-1.152&  -0.08622& -0.03483 & -0.05832  \\
$P_1$&-5.335   &-0.2024 &-3.690&  -0.04195&  0.07556&   0.05389 \\
$P_0$& 1.421   &3.606   &1.555 &  0.7323& 0.1073 &     0.6847\\
&&&&&&\\
\hline
\end{tabular}
\end{center}
\caption{\em Results for the double box graphs for Bhabha scattering}
\label{tablebhabha}
\end{table}
\normalsize

\subsection{Propagators up to 5 loops}

Due to the high complexity of multi-loop  QED and QCD calculations,  
only a few results at the three- and four loop 
level are known (for a review see e.g.\,\cite{Steinhauser:2002rq}). 
These results rely 
to a large extent on the knowledge of multi-loop propagator functions.

Massless propagator functions are very simple objects from the kinematical point
of view, as they depend only on one single scale, $s=p^2$, where $p$ is the
propagator momentum. Each graph is simply given by the scale to some power 
times a number
which can be calculated analytically or numerically once and forever.

\begin{figure}[ht]
\begin{picture}(150,30)
\put(5,22){(a)}
\put(10,0){\epsfig{file=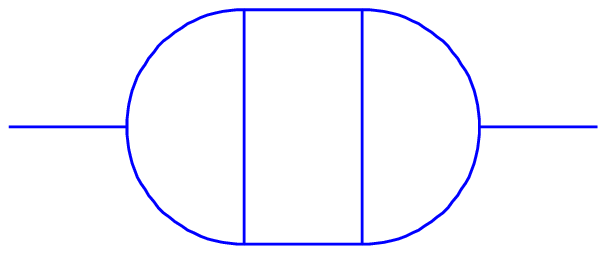,height=3.cm}}
\put(85,22){(b)}
\put(90,0){\epsfig{file=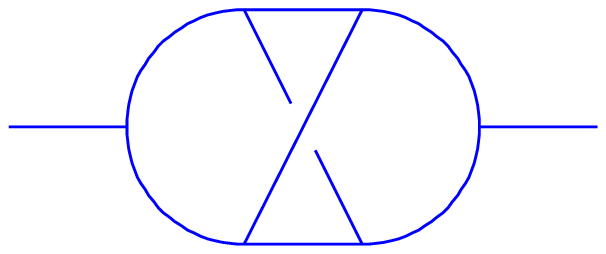,height=3.cm}}
\end{picture}
\caption{\em 3-loop ladder-type (a) planar and (b) non-planar  propagator graph.}\label{figTL3}
\end{figure}

\begin{figure}[ht]
\begin{picture}(150,30)
\put(5,22){(a)}
\put(10,0){\epsfig{file=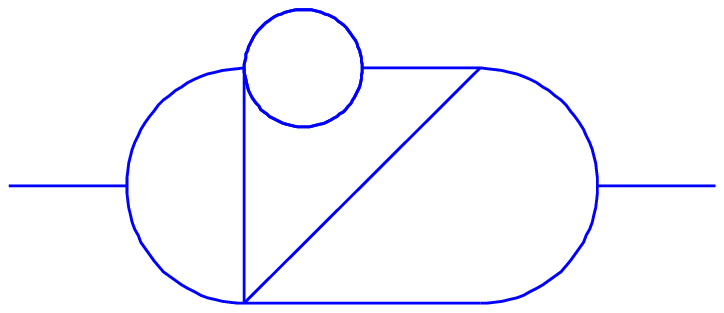,height=3.cm}}
\put(75,22){(b)}
\put(80,0){\epsfig{file=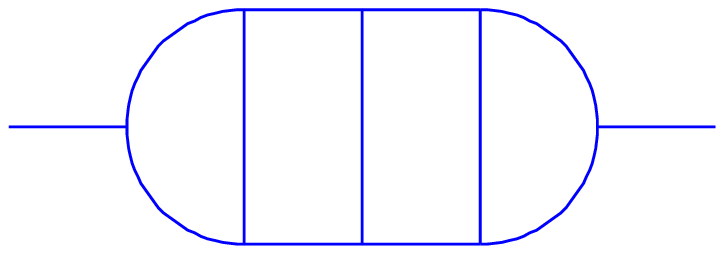,height=2.5cm}}
\end{picture}
\caption{\em 4-loop propagators (a) with and (b) without UV subdivergence.}
\label{figTL4}
\end{figure}

\begin{figure}[ht]
\begin{picture}(150,30)
\put(45,0){\epsfig{file=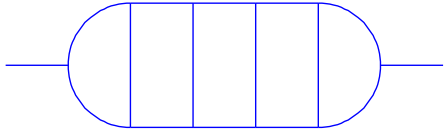,height=2.5cm}}
\end{picture}
\caption{\em A 5-loop propagator graph}\label{figTL5}
\end{figure}

We provide here some examples, shown in Figs.\,\ref{figTL3} to \ref{figTL5}, 
to demonstrate that our method can deal with
different kinds of propagator topologies up to 5 loops, and that the treatment
of UV subdivergences and higher order terms in the $\epsilon$-expansion 
is straightforward. 

\noindent
For the planar and non-planar  3-loop ladder (see Fig.\,\ref{figTL3}), we get:
\begin{eqnarray}
G[3{\rm a}]=(-s)^{-2-3\epsilon}\Gamma(2+3\epsilon)\left[20.74 + 93.71\,\epsilon \right] \\
G[3{\rm b}]=(-s)^{-2-3\epsilon}\Gamma(2+3\epsilon)\left[20.74 + 128.4\,\epsilon \right]
\end{eqnarray} 
For massless internal lines, UV subdivergences 
appear in the form of one-loop bubble insertions. 
The most convenient way to deal with them is to integrate them out 
analytically, which  leads to a non-integer exponent of an internal propagator.
An example is given in Fig.\,\ref{figTL4}a\,:
\begin{equation}
G[4{\rm a}]=-\frac{1}{\epsilon}(-s)^{-1-4\epsilon}\Gamma(1+4\epsilon)
{\rm Beta}(1-\epsilon,1-\epsilon)\,\left[ 20.74 + 86.57\,\epsilon+ 494.5\,\epsilon^2
\right]
\end{equation}
Note that in all three examples above the number 
 $20 \cdot \zeta(5) =   20.73855\dots $ appears with the
given precision\footnote{Going to higher precision is only a question
of computer time, as the integrands are positive definite and bounded.}.
\noindent Planar 4-loop ladder, Fig.\,\ref{figTL4}b\,:
\begin{equation}
G[4{\rm b}]=-(-s)^{-3-4\epsilon}\Gamma(3+4\epsilon)
\,\left[35.10+197.34\,\epsilon 
\right]
\end{equation}
The results for these propagator graphs agree with the analytical ones where 
available\,\cite{Chetyrkin}. Finally, we also calculated a 5-loop example, 
shown in Fig.\,\ref{figTL5}. We obtain: 
\begin{equation}
G[5]=(-s)^{-4-5\epsilon}\Gamma(4+5\epsilon)\,\left[ 40.53\right]
\end{equation}

\subsection{Massless on-shell planar 3-loop box}

\begin{figure}[ht]
\begin{picture}(150,30)
\put(45,0){\epsfig{file=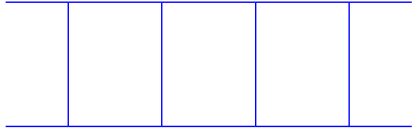,height=2.5cm}}
\end{picture}
\caption{\em Planar triple-box graph}\label{fig3box}
\end{figure}

In a very recent article\,\cite{Smirnov:new} (see also \cite{Smirnov:2002je}),  
Smirnov 
presented the analytical result for the massless on-shell planar 3-loop box
$TB(s,t)$, shown in Fig.\,\ref{fig3box}.  
We have cross-checked the analytical result by calculating
a numerical value for the point $(s,t)=(-1,-3)$, and obtain 

\begin{eqnarray*}
TB(-1,-3)&=&\Gamma(4+3\epsilon)\Bigl[
 \frac{0.09874}{\epsilon^6} 
-\frac{0.7669}{\epsilon^5} 
-\frac{1.977}{\epsilon^4} \nonumber\\
&&\hspace*{2.6cm}
-\frac{0.7534}{\epsilon^3} 
-\frac{4.747}{\epsilon^2} +\,\frac{2.010}{\epsilon} + 21.48\Bigr]
\end{eqnarray*}
which is in agreement within the 1\% level with the analytical result.
On a PC with a 2\,GHz Pentium IV processor, the full calculation took several weeks, 
but could have been speed up considerably  by further parallelisation. 
However, disk space starts to become an issue here, as the size of the  
object files adds up to about 500 Mega Bytes.

\section{Discussion and Conclusions}\label{discuss}

We have refined the formalism and automated program 
developed in \cite{Binoth:2000ps} to isolate overlapping
IR and UV divergences 
to be applicable to a very  large class of Feynman graphs. 
The algorithm now is able   
to treat diagrams with arbitrary masses and propagator exponents. 
As discussed above, this 
also allows to compute Feynman integrals with non-trivial numerators. 
Hence we have formulated a constructive approach, based on sector decomposition,
to convert dimensionally regulated Feynman diagrams into a Laurent series in $\epsilon$,  
where the coefficients are given in terms of parameter integrals.  
These parameter integrals can always be evaluated numerically 
in kinematic regions where the Mandelstam variables are negative, 
as one can show that the integrands are bounded and positive definite 
in that case.
As examples, we
calculated numerically various 2-loop 4-point functions and a 
3-loop 4-point function 
for negative Mandelstam variables $s,t,u$.  We also evaluated massless 
propagator graphs up to 5 loops.  
More precisely, we give numerical results for the following types of
diagrams:
\begin{itemize}
\item[--] Massless two-loop 4-point functions with two off-shell external legs (planar 
as well as non-planar topologies).
\item[--] The master two-loop box diagrams needed to calculate Bhabha scattering 
(with massive internal lines) at two loops. 
\item[--] Two-point functions at 3, 4 and 5 loops, among these a 4-loop graph 
containing an UV subdivergence, leading to non-integer propagator powers, 
as well as the order $\epsilon$ terms for the 3-loop and 4-loop graphs.
\item[--] The planar 3-loop massless 4-point function with on-shell legs.
\end{itemize}

Integrals depending on a single scale, as for example massless 2-point 
functions, or massless 
3-point functions with two on-shell legs, 
can be evaluated numerically if no analytical result is achievable,
as they are just a number times the overall scale factor. 
We note that the algorithm can also be applied to vacuum graphs with UV
subdivergences. In that case one has ${\cal F} = {\cal U} \sum_{j=1}^{N} x_j m_j^2$, 
and any graph can be expressed by well-behaved integrals. 
In this special situation other very powerful methods are known \cite{Laporta:2002pg}.   

For general multi-scale problems  
the situation is more delicate, as for physical kinematics
integrable singularities are present in a multi-dimensional integration
space. The numerical integration in this case is a highly non-trivial task. 
For the one-loop case
a combination of the sector decomposition algorithm and new numerical 
integration methods were proposed in \cite{Binoth:2002xh}, 
where it has been exploited that sector decomposition also leads to more 
stable integral representations if no IR/UV singularities are present.
A generalisation  to the two-loop case is presently under investigation.  
Once a procedure is set up to deal  with the sector integrals
numerically for general kinematics, a numerical approach for multi-leg and 
multi-loop processes relevant at present and future
high energy experiments will be feasible.
In the meantime, nonetheless, the algorithm serves as an independent  check for
analytical results of multi-loop integrals.
       
\section*{Acknowledgements}
G.H. would like to thank V.A.~Smirnov for a stimulating exchange of results.

\end{document}